\documentclass[aps,prb,twocolumn,longbibliography,preprintnumbers,amsmath,amssymb,superscriptaddress]{revtex4-1}

\usepackage{graphicx}
\usepackage[T1]{fontenc}
\usepackage[colorlinks,bookmarks=false,citecolor=blue,linkcolor=red,urlcolor=blue]{hyperref}
\usepackage[dvipsnames]{xcolor}
\usepackage{times}

\usepackage{bbm}

\newcommand{\be}{\begin{eqnarray}}
\newcommand{\ee}{\end{eqnarray}}

\newcommand{\nn}{\nonumber } 
\newcommand{\Eqref}[1]{Eq.~\eqref{#1}}

\begin{document}

\author{Daniel D. Scherer}\email{daniel.scherer@nbi.ku.dk}
\affiliation{Niels Bohr Institute, University of Copenhagen, DK-2100 Copenhagen, Denmark}

\author{Anthony Jacko}
\affiliation{School of Mathematics and Physics, The University of Queensland, Brisbane, Queensland, 4072, Australia}

\author{Christoph Friedrich}
\affiliation{Peter Gr\"{ü}nberg Institut and Institute for Advanced Simulation,
Forschungszentrum Jülich and JARA, 52425 Jülich, Germany}

\author{Ersoy \c{S}a\c{s}{\i}o\u{g}lu}
\affiliation{Peter Gr\"{ü}nberg Institut and Institute for Advanced Simulation,
Forschungszentrum Jülich and JARA, 52425 Jülich, Germany}
\affiliation{Institut f\"ur Physik, Martin-Luther-Universit\"at Halle-Wittenberg,  D-06099 Halle (Saale) Germany}

\author{Stefan Bl\"{u}gel}
\affiliation{Peter Gr\"{ü}nberg Institut and Institute for Advanced Simulation,
Forschungszentrum Jülich and JARA, 52425 Jülich, Germany}

\author{Roser Valent\'{i}}
\affiliation{Institut f\"{u}r Theoretische Physik, Goethe-University Frankfurt, 
Frankfurt am Main, Germany}

\author{Brian M. Andersen}
\affiliation{Niels Bohr Institute, University of Copenhagen, DK-2100 Copenhagen, Denmark}

\title{Interplay of nematic and magnetic orders in FeSe under pressure}

\begin{abstract}
\end{abstract}

\begin{abstract}
We offer an explanation for the recently observed pressure-induced magnetic state in the 
iron-chalcogenide FeSe based on \textit{ab initio} estimates for the pressure evolution of the most important Coulomb interaction parameters. We find that an increase of pressure leads to an overall decrease mostly in the nearest-neighbor Coulomb repulsion, which in turn leads to a reduction of the nematic order and the generation of magnetic stripe order. We treat the concomitant effects of band renormalization and the induced interplay of nematic and magnetic order in a self-consistent way and determine the generic topology of the temperature-pressure phase diagram, and find qualitative agreement with the experimentally determined phase diagram.
\end{abstract}

\maketitle

\section{Introduction}
\label{sec:intro}

The dominant electronic interactions that govern the low-energy physical
properties and the ordered phases of iron-based superconductors continue to
challenge the condensed matter community. In this respect recent intense
research efforts have focussed on the material FeSe due to its peculiar
properties. This material exhibits a prominent electronic driven (nematic)
structural phase transition setting in at $\sim90$ K below which the
$C_4$-symmetry of the lattice
is broken.
Importantly, for FeSe at ambient pressure there is no concomitant magnetic
transition at any lower temperatures in contrast to other known iron-based
superconductors.\cite{McQueen,Medvedev} For this reason nematic ordering
distinct from the spin-nematic scenario has been suggested.\cite{eremin,fernandes,christensen} FeSe is, however, poised to
magnetism \cite{glasbrenner2015} as evidenced by enhanced spin fluctuations,\cite{imai09,rahn,Zhao1,Zhao2},  and eventual
generation of static magnetic order at moderate uniaxial pressure above
$\sim1-2$ GPa.\cite{bendele10,bendele12,Kothapalli,sun16} The pressure-induced
magnetic order, which is known to be weak and to be consistent with stripe
order similar to the undoped magnetically ordered
compounds\cite{Kothapalli,khasanov,wang16}, emerges after the structural
transition (nematic phase) has been sufficiently suppressed by the
pressure.\cite{miyoshi,Terashima,sun16} Finally, the superconducting critical
temperature $T_c$ of FeSe is fascinatingly adjustable as seen both by its
approximately four-fold enhancement under
pressure,\cite{Medvedev,Kothapalli,miyoshi,Terashima,sun16} and by the
$T_c\sim100$ K for monolayer FeSe on STO substrates.\cite{wang12,Ge}

While the generation of nematic order in FeSe appears to be of electronic
origin,\cite{beak,fisher} the fundamental mechanism remains controversial
 at present.
Candidates include, for example, spontaneous orbital order as suggested by NMR
experiments\cite{beak,boehmer} and theoretical studies,\cite{onari,yamakawa}
frustrated magnetism,\cite{glasbrenner2015} quantum paramagnetism,\cite{lee} spin quadrupolar order,\cite{yu} or as a
result of competitive sub-leading charge-current density wave
order.\cite{chubukov} The open question of the origin of nematic order in FeSe
is presumably related to the sizable electronic interactions present in the
Fe chalcogenides.\cite{aichhorn,miyake,yin,evtushinsky2016,watson2016} Strong correlations may generate
distinct orbital selective properties for sufficiently large Hunds
coupling,\cite{ishida,si,medici,liu,yi15,backes2015} and such orbital selectivity seems
indeed present in FeSe as shown recently by a detailed modelling of the
superconducting gap anisotropy in this material.\cite{sprau,kreisel16}

Recently, yet another candidate was proposed for the origin of nematic order in
FeSe; longer ranged Coulomb interactions.\cite{Hu1,jiang2016} From {\it ab
initio} studies it is known that nearest-neighbor (NN) Coulomb repulsions are
larger for FeSe than in any of the other iron-based
superconductors,\cite{miyake} due to reduced screening from the lack of spacer
layers and/or the lower Fe-Fe bond lengths. Jiang {\it et al.}\cite{jiang2016}
and others~\cite{yi2015,wu2016} highlighted the importance of NN Coulomb
repulsions in FeSe, and showed that such longer-ranged interactions can both 1)
strongly renormalize the electronic structure and naturally generate small
Fermi pockets as seen in FeSe by ARPES and quantum
oscillations,\cite{nakayama,maletz,shimojima,terashima,watson1,zhang15,suzuki,fanfarillo,fedorov}
and 2) induce nematic site and bond order given by a spontaneous
splitting of the $d_{xz}$- and $d_{yz}$-dominant states. In
Ref.~\onlinecite{jiang2016} it was advocated that the competition of nematic
order with magnetic order also may explain the absence of magnetism in FeSe at
ambient pressure. 

Here, based on {\it ab initio} calculations for the pressure dependence of the
important interaction parameters including onsite $U$ and NN $V$ Coulomb
repulsions, we model the pressure dependence of both nematic and magnetic order
within the longer-range interaction scenario for nematic order described above.
We map out the general phase diagram of magnetic and nematic order and find
that a lowering of $V$ pushes the system from a purely nematic phase (driven by
$V$) into a magnetically ordered stripe phase (driven by $U$). As enhanced
pressure is found to decrease $V$ this offers a possible explanation of the
pressure-induced magnetic phase in FeSe. 
Finally we find also that the density of states near the Fermi level is larger in the magnetic phase than in the nematic phase, consistent with the overall increase of the superconducting $T_c$ with pressure.

We note that two recent theoretical studies also investigated
 the interplay of nematic and magnetic order in 
FeSe under pressure.\cite{glasbrenner2015,yamakawa_pressure}
In Ref.~\onlinecite{glasbrenner2015} a pressure dependent
 unusual magnetic frustration was identified via first principles calculations
while Ref.~\onlinecite{yamakawa_pressure} analyzed 
  the consequences for the
spin fluctuations of  a pressure-induced $d_{xy}$-dominant hole pocket.
Here we focus on the pressure-evolution of the interaction parameters
and pinpoint the important role of the NN Coulomb repulsion $V$ in explaining
the pressure-temperature phase diagram of FeSe.

The manuscript is structured as follows. We begin with a definition of the extended multiorbital Hubbard model in Sect.~\ref{sec:model} and briefly collect the set of self-consistent fields that enter the mean-field description of the correlated electronic system in Sect.~\ref{sec:self}. 
We then present our main results about the phase diagram of the model in Sect.~\ref{sec:tv} and provide a simple mechanism for the 
emergence of magnetism under application of pressure in FeSe. To connect the
parameter space of the extended multiorbital Hubbard model to the measured
pressure-temperature phase diagram, we analyze in Sect.~\ref{sec:dft} the
pressure dependence of hoppings and interaction parameters based on \textit{ab
initio} data. Finally, we discuss our results in Sect.~\ref{sec:conclusions}.
We collect details about the Hartree-Fock decoupling and the nematic order
parameter in Appendix~\ref{app:hf}. In Appendix~\ref{app:fs}, we summarize
known results about the band- and Fermi surface renormalization and nematic
order induced by strong NN Coulomb repulsion and provide  a RPA-level
instability analysis in the spin channel for the renormalized bandstructures to
show the enhanced spin density wave (SDW) ordering tendencies induced by NN Coulomb repulsion.

\section{Extended multiorbital Hubbard model}
\label{sec:model}

The itinerant electron system is described by a 5-orbital hopping Hamiltonian $H_{0}$ defined in the two-dimensional one-iron Brillouin zone (1-Fe BZ), a Hubbard-Hund interaction Hamiltonian $H_{U}$ and a NN Coulomb repulsion $H_{V}$,
\be
\label{eq:hamiltonian}
H = H_{0} + H_{U} + H_{V},
\ee
with
\be
\label{eq:hopping}
H_{0} = \sum_{\sigma}\sum_{i,j}\sum_{\mu,\nu} c_{i \mu \sigma}^{\dagger}\left( t_{ij}^{\mu\nu}  - \mu_{0} \delta_{ij}\delta_{\mu\nu} \right)c_{j \nu \sigma}, 
\ee
and
\be
\label{eq:interaction}
H_{U} & = &  
U \sum_{i,\mu} n_{i \mu \uparrow} n_{i \mu \downarrow} + 
\left(U^{\prime} - \frac{J}{2}\right) \sum_{i,\mu < \nu} n_{i \mu} n_{i \nu} \nn \\
& & \hspace{-2.5em} 
- 2 J \sum_{i, \mu < \nu}{\bf S}_{i\mu}\cdot{\bf S}_{i\nu}  + 
\frac{J^{\prime}}{2} \sum_{i, \mu \neq \nu,\sigma} c_{i\mu\sigma}^{\dagger}c_{i\mu\bar{\sigma}}^{\dagger}c_{i\nu\bar{\sigma}}c_{i\nu\sigma}.
\ee
as well as
\be
\label{eq:nncoul}
H_{V} = V \sum_{\langle i, j \rangle ,\mu,\nu} n_{i \mu} n_{j \nu}.
\ee
Here, the indices $\mu,\nu \in \{d_{xz}, d_{yz}, d_{x^2-y^2}, d_{xy}, d_{3z^{2}-r^{2}}\}$ specify the $3d$-Fe orbitals and $i,j$ run over the sites of the square lattice. The filling is fixed by the chemical potential $\mu_{0}$, and the onsite interaction is parametrized by an intraorbital Hubbard-$U$, an interorbital coupling $U^{\prime}$, Hund's coupling $J$ and pair hopping $J^{\prime}$. We will restrict ourselves to interaction parameters respecting orbital-rotational symmetry, which are realized for $U^{\prime} = U - J - J^{\prime}$, $J = J^{\prime}$. The fermionic operators $ c_{i \mu \sigma}^{\dagger}$, $c_{i \mu \sigma}$ create and destroy, respectively, an electron at site $i$ in orbital $\mu$ with spin polarization $\sigma$. Accordingly, we define the operators for local charge and spin as $n_{i\mu} = n_{i\mu\uparrow} + n_{i\mu\downarrow}$ with $n_{i\mu\sigma} = c_{i\mu\sigma}^{\dagger} c_{i\mu\sigma}$ and ${\bf S}_{i\mu} = \frac{1}{2}\sum_{\sigma\sigma^{\prime}} c_{i\mu\sigma}^{\dagger} {\boldsymbol \sigma}_{\sigma\sigma^{\prime}}c_{i\mu\sigma^{\prime}}$, respectively. Here, $ {\boldsymbol \sigma}$ denotes the vector of Pauli matrices. We specify the hopping parameters $t_{ij}^{\mu\nu}$ according to the bandstructure discussed in Ref.~\onlinecite{ningning2014} and neglect the effects of spin-orbit coupling.

\section{Self-consistent treatment of spin-density wave and bond-order mean-fields}
\label{sec:self}

We treat interaction effects in Hartree-Fock theory. To study the competition between stripe SDW order and nematic bond order, we decouple the onsite Hubbard-Hund term into the density fields
\be
\label{eq:mf1} 
n_{0}^{\mu\nu} = \frac{1}{\mathcal{N}} \sum_{{\bf k}, \sigma} 
\langle c_{{\bf k} \mu \sigma}^{\dagger} c_{{\bf k} \nu \sigma} \rangle,
\ee
and the magnetic order parameter
\be
\label{eq:mf2} 
M^{\mu\nu} = \frac{1}{\mathcal{N}} \sum_{{\bf k}, \sigma}\sigma 
\langle c_{{\bf k} + {\bf Q}, \mu \sigma}^{\dagger} c_{{\bf k} \nu \sigma} \rangle,
\ee
capturing the formation of collinear SDW order with ordering vector ${\bf Q} = (\pi,0)$ with antiferromagnetic
staggering of magnetization along $x$ between neighboring Fe sites and a ferromagnetic spin alignment along $y$. The ${\bf k}$ sum runs over the 1-Fe Brillouin zone and $\mathcal{N}$ denotes the number of unit cells. The density fields $n_{0}^{\mu\nu}$ describe orbital-dependent shifts and yield a weak renormalization of the Fermi surface.

For the NN Coulomb repulsion we adopt the Hartree-Fock decoupling into bond-order fields as introduced in Ref.~\onlinecite{jiang2016} to explain the band renormalization and nematic instability in FeSe. The self-consistent bond-order fields can be written as
\be 
\label{eq:mf3} 
\chi^{\mu\nu}({\bf k},\sigma) & = & \\
&  & \hspace{-1.5cm} \frac{1}{\mathcal{N}}\sum_{{\bf k}^{\prime}}\left[2\cos(k_{x} - k_{x}^{\prime}) + 2\cos(k_{y} - k_{y}^{\prime})\right]\langle c_{{\bf k}^{\prime} \nu \sigma}^{\dagger} c_{{\bf k}^{\prime} \mu \sigma} \rangle. \nn
\ee
The thermal average $\langle \cdots \rangle$ is computed with the eigenstates of the Bloch-Hamiltonian $h^{\mu\nu}({\bf k},\sigma)$ containing the mean-fields $n_{0}^{\mu\nu}$, $M^{\mu\nu}$, $\chi^{\mu\nu}({\bf k},\sigma)$. The Bloch-Hamiltonian is defined with respect to the reduced Brillouin zone $ [-\pi/2,\pi/2) \times [-\pi,\pi)$ and we decompose it according to the different self-consistent contributions as
\be
\label{eq:bloch}
h^{\mu\nu}({\bf k},\sigma) = h_{0}^{\mu\nu}({\bf k},\sigma) + h_{\mathrm{SDW}}^{\mu\nu}({\bf k},\sigma) + h_{\mathrm{BO}}^{\mu\nu}({\bf k},\sigma).
\ee
The bond-order field $\chi^{\mu\nu}({\bf k},\sigma)$ contains both $C_{4}$ symmetry-preserving and $C_{4}$ symmetry-breaking contributions that need to be treated separately. We refer to the former as the band renormalization (`$\mathrm{br}$') part, $\chi_{\mathrm{br}}^{\mu\nu}({\bf k},\sigma)$, while we denote the latter as the symmetry-breaking (`$\mathrm{sb}$') part $\chi_{\mathrm{sb}}^{\mu\nu}({\bf k},\sigma)$ serving as a nematic order parameter.~\cite{jiang2016} Accordingly, we introduce two different couplings $\tilde{V}$ and $\tilde{V}_{0}$ to control the effects of the $C_{4}$ symmetric ($\tilde{V}$) and $C_{4}$ breaking ($\tilde{V}_{0}$) contributions to the Hamiltonian on the electronic properties and replace $h_{\mathrm{BO}}^{\mu\nu}({\bf k},\sigma)$ by $\tilde{h}_{\mathrm{BO}}^{\mu\nu}({\bf k},\sigma)$ in Eq.~\ref{eq:bloch}, see Appendix~\ref{app:hf} for the explicit expression. The symmetry-preserving part was shown to yield a substantial band renormalization~\cite{jiang2016} emerging in a more or less natural way from repulsive NN interactions. With properly chosen $\tilde{V}$, the electronic band structure is prone to a nematic instability triggered by $\tilde{V}_{0} \neq 0$. We note, that for $V_{0} = 0$, no nematic instability can occur within our mean-field approach. Since $\tilde{V}$ and $\tilde{V}_{0}$ are the couplings of operators transforming differently under point group operations, it is natural to assume $\tilde{V} \neq \tilde{V}_{0}$ can occur by applying a renormalization group procedure to high-energy degrees of freedom. Since we are not attempting a quantitative determination of these renormalization processes or the corresponding couplings from the microscopic interaction parameters, we denote the phenomenological couplings by a tilde to distinguish them from the bare microscopic NN Coulomb interaction. Below we use $\tilde{V}_{0} > \tilde{V}$ to generate a moderate amplitude for the nematic order parameter. We expand $ \chi^{\mu\nu}({\bf k},\sigma) $ in NN form factors $f_{\mathrm{A}}({\bf k})$ as
\be
\chi_{\mathrm{br}/\mathrm{sb}}^{\mu\nu}({\bf k},\sigma) = \sum_{A}\chi_{\mathrm{br}/\mathrm{sb},A}^{\mu\nu}(\sigma)f_{A}({\bf k}), \, A = s, p_{x}, p_{y}, d.
\ee
We then solve the set of self-consistent equations numerically to determine the mutual influence of band renormalization,
nematic and SDW order. Details on the self-consistent mean-field approach are collected in Appendix~\ref{app:hf}.
%
\begin{figure}[t!]
\includegraphics[width=1\columnwidth]{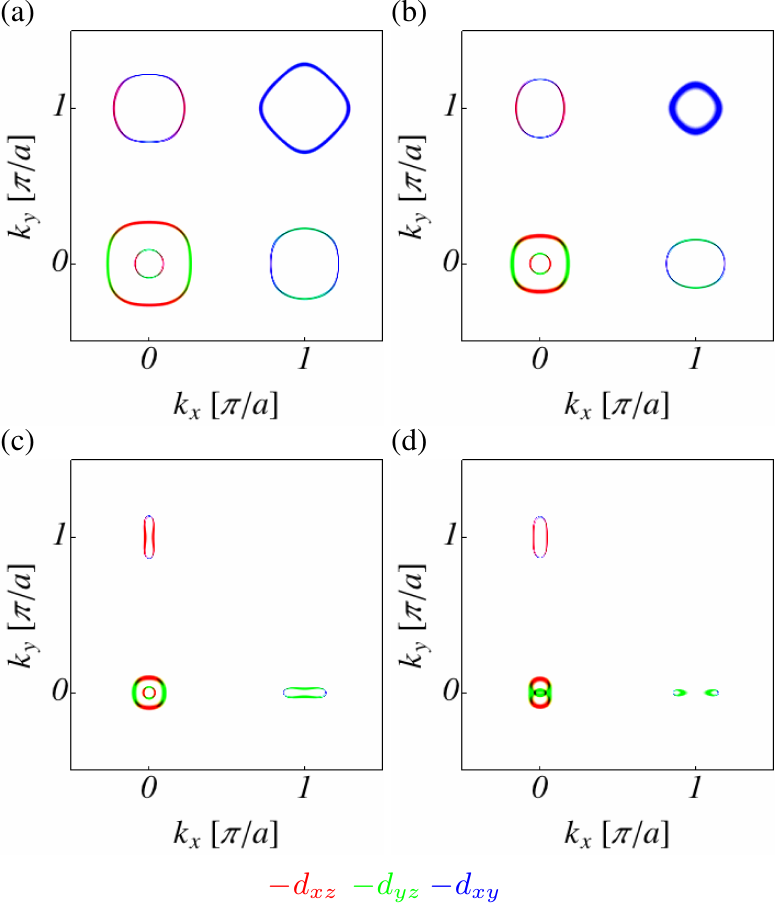}
\caption{(a) - (d) Orbital resolved Fermi surfaces obtained from the spectral function at filling $n = 6$ for different values of the NN Coulomb interaction parameters. Renormalizations due to the local interaction are typically small and are neglected here. (a) The Fermi surface of the tight-binding model without additional renormalizations, $\tilde{V} = \tilde{V}_{0} = 0$. (b,c) The Fermi surface including the self-consistent band renormalization $\chi_{br}^{\mu\nu}$ due to NN Coulomb repulsion, (b) $\tilde{V} = 0.35 \, \mathrm{eV}$, (c) $\tilde{V} = 0.74 \, \mathrm{eV}$. (d) The Fermi surface in the nematic state stabilized by Fermi surface renormalization and a self-consistently $C_{4}$ symmetry-breaking contribution, $\tilde{V} = 0.74 \, \mathrm{eV}$, $\tilde{V}_{0}/\tilde{V}= 1.8 $.}
\label{fig:fs}
\end{figure}
%
We also introduce the nematic order parameter in the $d$-wave channel as
\be 
\Delta_{d} = \frac{1}{2}\sum_{\sigma}(\chi_{\mathrm{sb,d}}^{xz,xz}(\sigma) + \chi_{\mathrm{sb,d}}^{yz,yz}(\sigma)),
\ee
that was established as the leading bond-order wave component in the nematic state triggered by the van Hove singularity for finite $\tilde{V}_{0}$.~\cite{jiang2016} The dramatic band renormalization and the deformation of the Fermi surface through nematic order are demonstrated in Fig.~\ref{fig:fs} and found to be very similar to the recently obtained Fermi surface as extracted from e.g. quasi-particle interference.\cite{sprau} More details on the band renormalization can also be found in Appendix~\ref{app:fs}.

\section{temperature-pressure phase diagram}
\label{sec:tv}
%
\begin{figure*}[t!]
\begin{minipage}{1\textwidth}
\centering
\includegraphics[width=1\columnwidth]{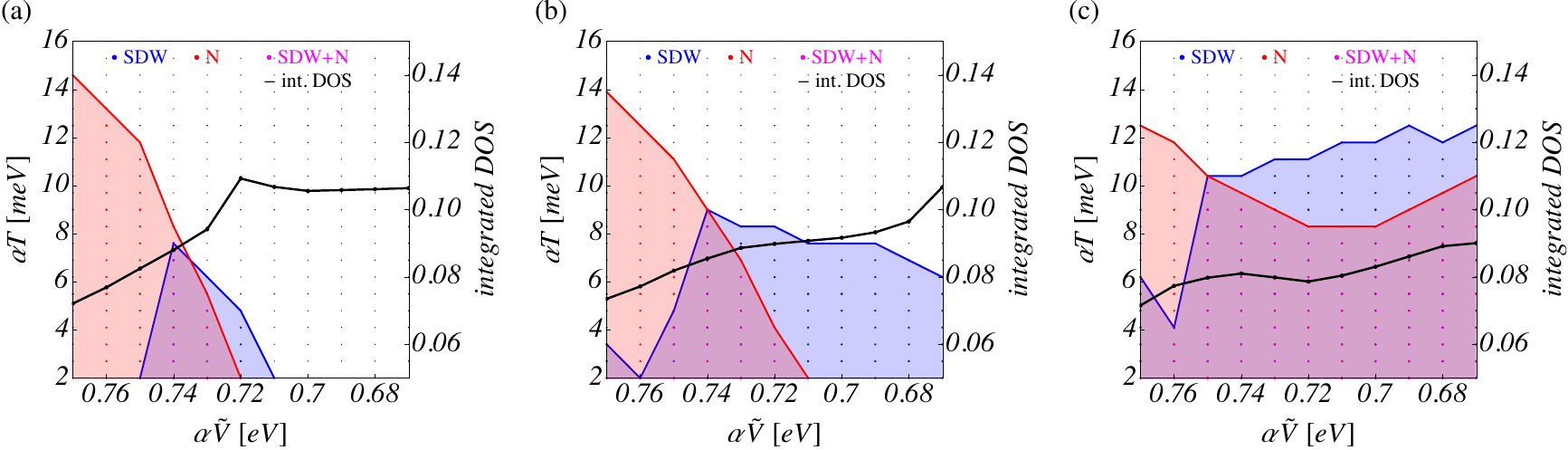}
\end{minipage}
\caption{(a) - (c) Phase diagrams in NN Coulomb interaction strength $ \alpha \tilde{V} $ vs. temperature $ T / \alpha $ parameter space for the FeSe model. The red shaded region denotes the nematic state (N), while the blue shaded region corresponds to the magnetic stripe state (SDW). The coexistence of nematic order and magnetic order is indicated by magenta color. The points indicate the parameters where self-consistent calculations have been carried out. We fixed the ratio of $\tilde{V}_{0}/\tilde{V} = 1.8$ generating a splitting of about $50$ meV of the degenerate $d_{xz}$ and $d_{yz}$ orbitals at the $M$ point (with respect to the 2-Fe BZ) in the nematic state. For the displayed phase diagrams we fix $\alpha U = 1.40$ eV and study the impact of the rescaled Hund's coupling $\alpha J$ on the phase diagram, where in (a) $ \alpha J  = 0.325 $ eV, (b) $ \alpha J = 0.350 $ eV and (c) $ \alpha J = 0.375 $ eV. The black curves show the integrated density of states in an energy range $[-25,+25]$ meV to around the Fermi level with $ \alpha T = 2 $ meV as a function of $ V $ to obtain a naive estimate of the superconducting $T_{\mathrm{c}}$.}
\label{fig:pd}
\end{figure*}
In the following, we will present our main result, namely the $ \tilde{V} - T $ phase diagram for a multiorbital model 
of FeSe, where we allow for stripe SDW order with ordering vector ${\bf Q} = (\pi,0)$ and uniform nematic bond order. We note, that we take the full order parameter $\chi_{\mathrm{br}}^{\mu\nu}$ into account, without restricting to the $d_{xz}, d_{yz}$ subspace.

In modeling the pressure effects, based on \textit{ab initio} findings discussed  in Sect.~\ref{sec:dft}, we proceed in the following way. We take the Hamiltonian \Eqref{eq:hamiltonian} and assume a uniform renormalization of the hopping matrix elements under application of pressure and therefore replace $t_{ij}^{\mu\nu} \to \alpha^{-1}(p) t_{ij}^{\mu\nu} $ in the kinetic term $H_{0}$ with a renormalization factor $\alpha^{-1}(p) \geq 1$, where we made the dependence of pressure $p$ explicit.
Thus the Hamiltonian is written as
\be 
H = \alpha^{-1}(p) H_{0} + H_{U(p)} + H_{V(p)},
\ee
where we also replaced the couplings of the onsite interactions by pressure dependent functions. To work with a fixed bandstructure we rescale the Hamiltonian by the factor $\alpha(p)$ and arrive at a rescaled Hamiltonian 
\be 
H^{\prime}(p) = \alpha(p) H = H_{0} + H_{\alpha(p) U(p)} + H_{\alpha(p) V(p)}.
\ee
We note that such a rescaling in principle also entails a rescaling of temperature. Since our aim in this
work is to provide a proof of principle that the NN Coulomb repulsion is the relevant variable
that is responsible for the topology of the temperature-pressure phase diagram, we refrain from
determining an approximation for $\alpha(p)$ and instead determine a
phase diagram in the parameter space spanned by $\alpha T$ and $\alpha
\tilde{V}$. As an additional simplification, we take $\alpha(p) U(p)$ and
$\alpha(p) J(p)$ as well as the other local couplings to be constant. The
subsequent mean-field decoupling and splitting of the rescaled coupling $\alpha
V$ into $\alpha \tilde{V}$ and $\alpha \tilde{V}_{0}$ proceeds as explained in
Sect.~\ref{sec:self}. For concreteness, we fix the ratio $ \tilde{V}_{0} /
\tilde{V} = 1.8 $. The value of $\tilde{V}_{0}$ essentially controls the size
of the nematic $d$-wave order parameter $\Delta_{d}$ and thereby the size of
the splitting between $d_{xz}$ and $d_{yz}$ orbitals at high-symmetry points in
the BZ. The splitting at the $M$ point in the 2-Fe BZ is about $50$ meV in the
low-temperature nematic phase for this choice of parameters. While this value
might overestimate the size with respect to the experimentally observed
spectral splitting~\cite{watson20162}, we note that on the level of our self-consistent mean-field
description, fluctuation effects of nematic and magnetic order parameters are
not included. We expect that including their feedback on the phase diagram will
lead to a downward renormalization of critical temperatures and the magnitudes
of the order parameters. Changing the value of $\tilde{V}_{0}$ for fixed $\tilde{V}$ effectively tunes both the size of this splitting and the extent of the nematic phase.~\cite{jiang2016} The topology of the phase diagram remains robust, however, to changing the ratio $\tilde{V}_{0}/\tilde{V}$.

Solving the self-consistent mean-field equations yields the phase diagrams shown in Fig.~\ref{fig:pd}(a)-\ref{fig:pd}(c), where we used an $ 80 \times 80 $ grid-discretization of the 1-Fe BZ in the numerical implementation. We consider the system in a nematic (magnetic) state if the nematic order parameter (magnetic moment) exceeds a numerical value of $ 5 \times 10^{-3} $. Otherwise we consider the system to be in a paramagnetic state. We note that we reversed the $ \alpha \tilde{V} $ axis in our phase diagrams, such that decreasing $ \alpha \tilde{V} $ corresponds to increasing pressure, in order to facilitate an easier comparison to the experimental phase diagrams. We restrict our attention to the interval $ \alpha \tilde{V} \in [ 0.67, 0.77 ] $ eV, corresponding to a pressure-induced decrease of the rescaled coupling $ \alpha \tilde{V} $ by $ \sim 13 $ percent, which we here take as a conservative guess of the true order of magnitude of the pressure effects on NN Coulomb repulsion.

The onsite intra-orbital repulsion was taken to be $ \alpha U = 1.40 $ eV. For $ \alpha J < 0.325$ eV, we observe no magnetic order in the $ \alpha \tilde{V} $ range we consider in Fig.~\ref{fig:pd}. We study the influence of the Hund's coupling on the phase diagram by looking at the cases $ \alpha J = 0.325, 0.350, 0.375 $ eV. In line with an RPA-instability analysis, 
see Appendix~\ref{app:fs}, the stripe SDW order sets in around $ \alpha
\tilde{V} = 0.73 $ eV and forms a little dome at the foot of the nematic phase.
The nematic phase can of course be stabilized for vanishing onsite interactions
and is completely driven by band renormalization due to $ \alpha \tilde{V} $
and the coupling $\alpha \tilde{V}_{0}$ triggering the nematic symmetry breaking. 
We performed the same mean-field analysis for the interaction parameters $ \alpha U = 1.30, 1.50 $ eV and $ \alpha J = 0.325, 0.350, 0.375 $ eV (not shown in Fig.~\ref{fig:pd}). As expected, decreasing $\alpha U$ reduces the SDW ordering tendencies, while increasing $\alpha U$ boosts SDW order (and correspondingly the size of the ordered magnetic moment). The phase diagrams in Fig.~\ref{fig:pd} are representative in the sense that they already capture the main trends.

The magnetic and the nematic phase show little `competition' effects: most of the
SDW dome coexists with the nematic phase. The extent of the SDW phase increases
as the Hund's coupling grows. At the same time, the presence of a finite SDW
order parameter sources a finite nematic order parameter, as can be expected
from symmetry considerations. Both SDW and nematic phase break the $C_{4}$
symmetry of the lattice, while the SDW also breaks SU(2)-spin and time-reversal
symmetry. By formally expanding the mean-field free-energy in $M^{\mu\nu}$ and
$\chi_{\mathrm{sb}}^{\mu\nu}$ one obtains a coupling of the modulus of the SDW
order parameter to the nematic bond-order parameter. Additionally, the two
types of instabilities are driven by different microscopic interactions. 

Nevertheless, we emphasize that the two different orders are intertwined in the considered $\tilde{V}$ range for the following reasons: i) the band renormalization due to NN Coulomb pushes the van-Hove singularity close to the Fermi level and ii) optimizes $(\pi,0) / (0,\pi)$ nesting of the central hole pocket and the electron pockets. While i) enables the formation of the nematic state, it is ii) that gives rise to an SDW dome of finite extent for values of the Hund's coupling that is large enough to trigger a SDW instability but not large enough to cross the SDW threshold also for the non-optimally nested cases. Previous theoretical studies\cite{mukherjee,kreisel} modelling NMR\cite{beak,imai09} and neutron scattering\cite{rahn,Zhao1,Zhao2} also concluded that FeSe is close to a magnetic instability.
\begin{figure}[t]
\begin{minipage}{1\columnwidth}
\centering
\includegraphics[width=1\columnwidth]{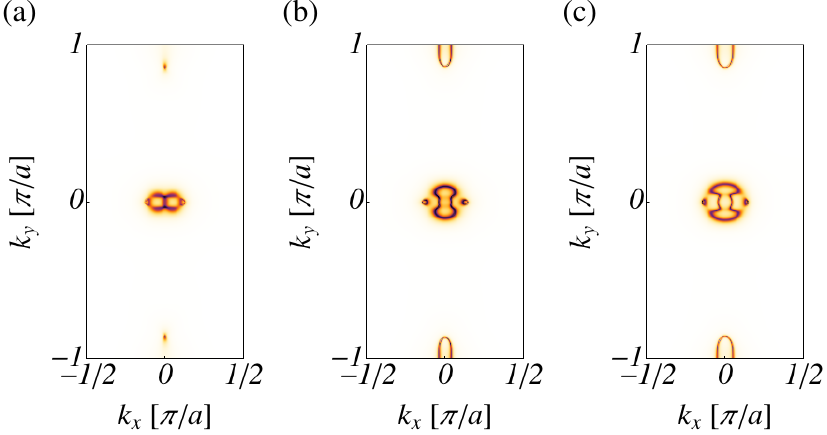}
\end{minipage}
\caption{(a) - (c) Reconstructed Fermi surfaces for (a) $ \alpha \tilde{V} = 0.77 $ eV, (b) $ \alpha \tilde{V} = 0.73 $ eV and (c)
$ \alpha \tilde{V} = 0.69 $ eV for $ \alpha U = 1.40 $ eV and $ \alpha J = 0.350$ eV at $ \alpha T = 2 $ meV, corresponding to the
states in the phase diagram shown in Fig.~\ref{fig:pd}(b).}
\label{fig:refs}
\end{figure}
Within our mean-field description the ordered magnetic moment depends sensitively on temperature. At the lowest temperatures, we obtain ordered moments ranging from $ 0.04 - 0.12 \mu_{\mathrm{B}}$, which is roughly in agreement with the experimentally reported values in the magnetic phase of FeSe~\cite{khasanov} for the largest ordered moments we found. Naturally, the largest value of 
$ 0.12 \mu_{\mathrm{B}} $ for the ordered moment is realized for larger Hund's coupling, here $ \alpha J = 0.375 $ eV. 

In order to complete the phase diagrams Fig.~\ref{fig:pd}(a)-\ref{fig:pd}(c) we also need an estimate of the evolution of the superconducting $T_{\mathrm{c}}$. We leave the determination of the fluctuation induced Cooper vertex and the solution of the corresponding gap equation for future work and restrict ourselves to a `poor man's argument' by examining the $\tilde{V}$ dependence of the integrated density of states (DOS), see Fig.~\ref{fig:pd}(d)-Fig.~\ref{fig:pd}(f). We chose a symmetric integration interval of width $ 50 $ meV around the Fermi level. We observe that as the size of the nematic order parameter decreases the integrated DOS tends to increase. This observation remains true in the SDW-dominated regime. If we now take the integrated DOS as a proxy for the system's tendency to build up a superconducting condensate,
it is likely that an increase of $T_{\mathrm{c}}$ with decreasing $ \alpha \tilde{V} $ can be observed, in agreement with experiment. Finally, we show representative reconstructed Fermi surfaces in Fig.~\ref{fig:refs}(a)-(c) for parameters relevant to panel (b) of Fig.~\ref{fig:pd}. As seen the reconstructed bands contain new tiny Fermi pockets which seem in overall agreement with recent quantum oscillations measurements of FeSe under pressure.\cite{Terashima_recon}

\section{Pressure-induced renormalization of bandstructure and couplings from DFT}
\label{sec:dft}

In this section we want to connect the band and interaction parameters, particularly 
the NN Coulomb interaction, to the application of hydrostatic pressure
on FeSe. We extract these parameters from \textit{ab initio} calculations
on FeSe crystal structures for the  pressure range from $0 - 10$ GPa~\cite{Milan}.
\begin{figure}[h!]
\begin{minipage}{0.95\columnwidth}
\centering
\includegraphics[width=1\columnwidth]{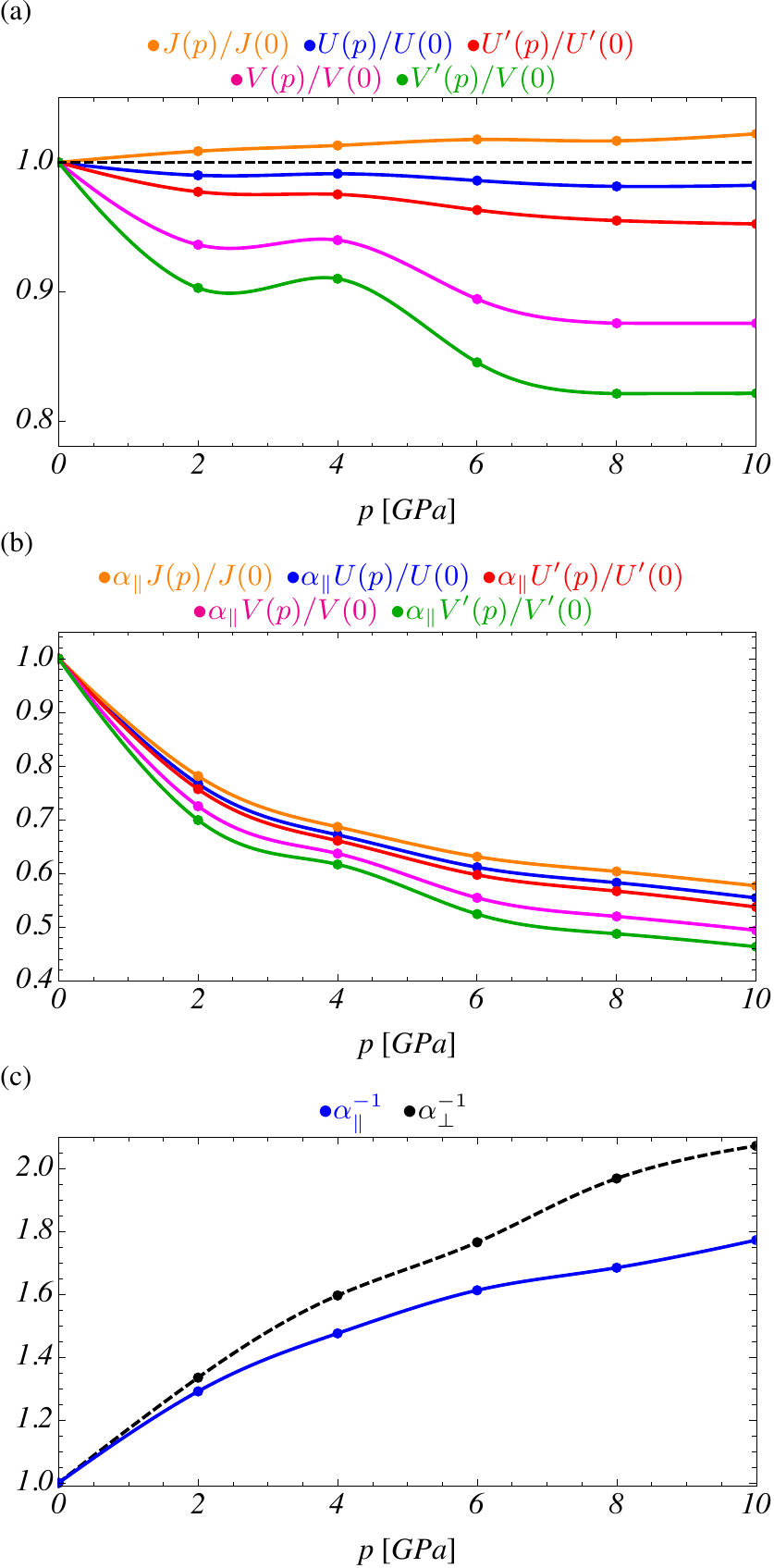}
\end{minipage}
\caption{(a) Pressure-induced renormalization of onsite ($U, U^{\prime}, J$), NN ($V$) and NNN ($V^{\prime}$) couplings extracted from DFT calculations relative to the $p = 0$ MPa value. The only coupling that shows a clear upward trend under application of pressure is Hund's coupling $J$. The intra- and interorbital repulsions $U$ and $U^{\prime}$ as well as the longer-ranged repulsion $V$ and $V^{\prime}$ decrease with pressure. The longer-ranged interactions are clearly more affected and display changes between $\sim 10$\,\% ($V$) and $\sim 15$\,\% ($V^{\prime}$). (b) Pressure dependence of couplings rescaled with a rough estimate of the hopping renormalization from intra-layer hoppings, $\alpha_{\parallel}^{-1}$. (c) Estimates of the hopping renormalizations for inter- and intra-layer hoppings, $\alpha_{\perp}^{-1}$ and $\alpha_{\parallel}^{-1}$ as a function of pressure.}
\label{fig:couplings}
\end{figure}
We used the FLEUR package, a full-potential linearized augmented-plane-wave
(FLAPW) density functional theory method to compute the ground state
density,\cite{flapw} and the Spex code~\cite{friedrich}
 to perform constrained RPA (cRPA) calculations~\cite{} to find
the screened Coulomb interactions. Densities were converged on
an $8 \times 8 \times 8$ k-mesh with a Perdew-Burke-Ernzerhof 
non-relativistic functional.\cite{perdew} The active space of the cRPA calculation was the five
$3d$ orbitals per Fe.
The tight-binding parameters were obtained~\cite{Tomic_unfolding,Guterding_parameters} 
by using projective Wannier functions
as implemented in the
all electron full potential local orbital (FPLO) code.~\cite{kopernik}

We show the effect of pressure on the onsite and longer-ranged couplings in Fig.~\ref{fig:couplings} that are obtained as orbital averages of orbital resolved interaction matrices extracted from the DFT calculations. Here, we denote the next-nearest-neighbor (NNN) repulsion by $V^{\prime}$. Interestingly, the couplings $U(p), U^{\prime}(p), V(p), V^{\prime}(p)$ show a downward trend under increasing pressure, see Fig.~\ref{fig:couplings}(a), while only the Hund's coupling $J(p)$ slightly increases. At intermediate pressures, the longer-ranged couplings $V, V^{\prime}$ show non-monotonous behavior. In this work, we refrain from performing our instability analysis for the bandstructures obtained for 
different hydrostatic pressures. Instead, we focus on the most dominant trends. To simplify our calculations, we
keep the bandstructure fixed and only renormalize the couplings. To estimate the changes in the 3D bandstructures obtained from the present DFT calculations in a semi-quantitative way, we arrange all hopping matrix elements $t_{ij}^{\mu\nu}(p)$ for a given pressure in a vector ${\bf t }$ and compute the Euclidean norm $ || {\bf t }(p) || $. We then define the renormalization factors $ \alpha_{\parallel}^{-1}(p) = || {\bf t }_{\parallel}(p) || / || {\bf t }_{\parallel}(0) || $ and $ \alpha_{\perp}^{-1}(p) = || {\bf t }_{\perp}(p) || / || {\bf t }_{\perp}(0) || $ for in- and out-of-plane hoppings. Both $\alpha_{\parallel}^{-1}(p)$ and $\alpha_{\perp}^{-1}(p)$ show an upward trend as the pressure is increased, see Fig.~\ref{fig:couplings}(c). We note that these renormalizations give only a gross estimate of the effect of pressure on the bandstructure, and additionally the precise values and even the ratio of in-plane to out-of-plane renormalization also depend on the choice of the norm.

The upward evolution, however, is a robust feature. To estimate the evolution of the couplings relative to the increase in bandwidth in a 2D system, we rescale the couplings of Fig.~\ref{fig:couplings}(a) by the factor $ \alpha_{\parallel}(p) $ and show the pressure evolution in Fig.~\ref{fig:couplings}(b). The rescaled couplings all show a decrease with increasing pressure. The effect on the longer-ranged couplings is in any case dominating.

Therefore we suggest that the leading effect of pressure on the nematic and magnetic orders can be obtained from the decrease in the NN Coulomb repulsion, leading to the phase diagram presented in the previous section.

\section{Conclusions \& Discussion}
\label{sec:conclusions}

In this work we have studied the interplay of nematic bond order and
stripe magnetism in an extended multiorbital Hubbard model for FeSe.
We propose an explanation for the experimentally 
observed temperatue-pressure phase diagram of FeSe in terms of a pressure-induced decrease of the 
NN Coulomb interaction. Assuming that the formation of the nematic phase is 
driven by a band renormalization due to NN Coulomb, where the size of the magnetic
order parameter is controlled by a coupling in a different symmetry channel, the decrease of the
Coulomb repulsion moves the relevant van Hove singularity away from the Fermi
level and at the same time optimizes the nesting condition for stripe magnetism.
This naturally explains the decrease of nematic order and the 
emergence of magnetic order under the application of pressure.
Concerning the superconducting properties of FeSe, we attempted to provide
a crude estimate for the ordering tendencies based on the integrated density of states
around the Fermi level, which displays an increase for decreasing the NN Coulomb
interaction. 

In our modeling of the pressure dependence of hoppings and interaction parameters,
we were guided by the results of \textit{ab initio} calculations taking the effect of
pressure into account. From these results we distilled a simplified model assuming
that pressure influences all hoppings uniformly and can thus be treated by a global
rescaling of hopping parameters. Interestingly, the \textit{ab initio} results suggest
that both onsite and longer-ranged couplings decrease under application of pressure, 
with the exception of the Hund's coupling which displays a slight
increase. The longer-ranged interactions show the largest decrease.
This pressure dependent decrease of the interaction parameters is attributed
to an increase of the effective screening as the nuclei come closer.

Within our model, the important ingredient
is a decrease of the \textit{rescaled} NN Coulomb interaction as a function of pressure. 
Our conclusions are therefore robust, as long as the renormalization of the bandwidth
due to pressure overcompensates a possible increase of NN Coulomb under pressure,
as one might naively expect.

We also note that taking a rescaling of temperature and onsite interactions into account
does not change the main conclusions about the topology of the phase diagram and the underlying mechanism, as long as the initial values $U(0), J(0)$ of the onsite couplings at pressure $ p  = 0 $ are chosen appropriately.

\begin{acknowledgments}

We thank P. J. Hirschfeld, A. Kreisel, M. Tomi{\'c} and D. Guterding for useful discussions. D.D.S. acknowledges support from the Villum foundation. A. J. acknowledges support through the Australian Research Council grant DP160100060. S.B. and R.V. achnowledge support through the Deutsche Forschungsgemeinschaft (DFG) through grants SPP1458 (R.V.) and FOR 1346 (S.B. and R.V.). B.M.A acknowledges support from Lundbeckfond fellowship (grant A9318).

\end{acknowledgments}

\newpage


\begin{widetext}


\appendix


\section{Hartree-Fock decoupling and nematic order sparameter}
\label{app:hf}

We treat interaction effects in Hartree-Fock theory. To study the competition between stripe SDW order and nematic bond order, we decouple the onsite Hubbard-Hund term into the fields
\be
\label{eq:app:mf1} 
n_{0}^{\mu\nu} = \frac{1}{\mathcal{N}} \sum_{{\bf k}, \sigma} 
\langle c_{{\bf k} \mu \sigma}^{\dagger} c_{{\bf k} \nu \sigma} \rangle,
\quad
M^{\mu\nu} = \frac{1}{\mathcal{N}} \sum_{{\bf k}, \sigma}\sigma 
\langle c_{{\bf k} + {\bf Q}, \mu \sigma}^{\dagger} c_{{\bf k} \nu \sigma} \rangle,
\ee
with ${\bf Q } = (\pi,0)$, while the NN Coulomb repulsion is decoupled into bond-order order parameters as
\be 
\label{eq:app:mf2} 
\chi^{\mu\nu}({\bf k},\sigma) = \frac{1}{\mathcal{N}}\sum_{{\bf k}}\left[2\cos(k_{x} - k_{x}^{\prime}) + 2\cos(k_{y} - k_{y}^{\prime})\right]\langle c_{{\bf k}^{\prime} \nu \sigma}^{\dagger} c_{{\bf k}^{\prime} \mu \sigma} \rangle
\ee
The average $\langle \cdots \rangle$ on the right hand side is computed with respect to a thermal state of the Hartree-Fock Hamiltonian $H_{\mathrm{HF}} = \sum_{{\bf k},\mu,\nu\sigma}^{\prime}\Psi_{{\bf k}\mu\sigma}^{\dagger}h^{\mu\nu}({\bf k},\sigma)\Psi_ {{\bf k}\nu\sigma}$. The Bloch-Hamiltonian $h^{\mu\nu}({\bf k},\sigma)$ containing the mean-fields Eqs. (\ref{eq:app:mf1})-(\ref{eq:app:mf2}) is defined with respect to the reduced Brillouin zone $ [-\pi/2,\pi/2) \times [-\pi,\pi)$. We decompose it as
\be 
h^{\mu\nu}({\bf k},\sigma) = h_{0}^{\mu\nu}({\bf k},\sigma) + h_{\mathrm{SDW}}^{\mu\nu}({\bf k},\sigma) + h_{\mathrm{BO}}^{\mu\nu}({\bf k},\sigma),
\ee
where
\begin{eqnarray}
h_{0}^{\mu\nu}({\bf k},\sigma) & = & 
\begin{pmatrix}
 \xi^{\mu\nu}({\bf k}) + N^{\mu\nu}_0   & 0 \\
0 & \xi^{\mu\nu}({\bf k}+{\bf Q}) + N^{\mu\nu}_0 \\
\end{pmatrix}, \\
h_{\mathrm{SDW}}^{\mu\nu}({\bf k},\sigma) & = & 
\begin{pmatrix}
0  & \sigma W^{\mu\nu}  \\
 \sigma W^{\mu\nu} & 0 \\
\end{pmatrix}, \\
h_{\mathrm{BO}}^{\mu\nu}({\bf k},\sigma) & = & 
-\frac{V}{2}
\begin{pmatrix}
 \chi^{\mu\nu}({\bf k},\sigma) & 0  \\
0 & \chi^{\mu\nu}({\bf k}+{\bf Q},\sigma) \\
\end{pmatrix} + (\mu \leftrightarrow \nu)^{\ast}.
\end{eqnarray}
The basis is defined by the spinor
\begin{eqnarray}
\Psi_{{\bf k}\mu\sigma}^{\dagger}=
\begin{pmatrix}
c_{{\bf k}\mu\sigma}^{\dagger} &
c_{{\bf k} + {\bf Q}\mu\sigma}^{\dagger}
 \end{pmatrix},
 \,\,\,
\Psi_{{\bf k}\mu\sigma}=
\begin{pmatrix}
c_{{\bf k}\mu\sigma} \\
c_{{\bf k} + {\bf Q}\mu\sigma}
 \end{pmatrix}, 
\end{eqnarray}
and the mean fields Eq. (\ref{eq:app:mf1}) enter through the quantities
\be
\label{eq:app:N0}
N_{0}^{\mu\nu} & = & 
\delta^{\mu\nu}\Bigl(U n_{0}^{\mu} + (2 U^{\prime} - J) \bar{n}_{0}^{\nu}\Bigr) + \bar{\delta}^{\mu\nu} \Bigl((-U^{\prime} + 2 J) n_{0}^{\nu\mu} + J^{\prime} n_{0}^{\mu\nu} \Bigr),
\ee
and
\be
\label{eq:app:W1}
W^{\mu\nu} & = & \delta^{\mu\nu}\Bigl(-U M^{\mu} - J \bar{M}^{\nu} \Bigr) + \bar{\delta}^{\mu\nu}\Bigl( U^{\prime} M^{\nu\mu} - J^{\prime} M^{\mu\nu} \Bigr).
\ee
Here, $\bar{\delta}^{\mu\nu} = 1 - \delta^{\mu\nu}$ filters out the orbital off-diagonal components. We note, that repeated indices are \emph{not} summed over in the above expressions. Quantities in Eqs.~(\ref{eq:app:N0}),(\ref{eq:app:W1}) with a single orbital index refer to the diagonal element of the corresponding matrix, e.g. $ n_{0}^{\mu} = n_{0}^{\mu\mu} $. Objects with a bar, such as $\bar{n}_{0}^{\nu}$, are defined as, e.g., $\bar{n}_{0}^{\nu} = \sum_{\mu \neq \nu} n_{0}^{\mu\mu}$. The bare dispersion enters through $\xi^{\mu\nu}({\bf k}) = \varepsilon^{\mu\nu}({\bf k}) -   \delta^{\mu\nu} \mu_{0}$, where $\epsilon^{\mu\nu}({\bf k}) $ is obtained from the Bloch representation of the hopping Hamiltonian \Eqref{eq:hopping} and $\mu_{0} $ is the chemical potential controlling the filling of the electronic bands.

The bond-order Hamiltonian needs to be treated with care. As demonstrated in Ref.~\onlinecite{jiang2016} the bond-order mean-field $\chi^{\mu\nu}({\bf k},\sigma)$ contains both $C_{4}$ symmetry-preserving and $C_{4}$ symmetry-breaking contributions that need to be treated separately. The symmetry-preserving part was shown to yield a substantial band renormalization, emerging in a more or less natural way from including repulsive NN interactions. Obviously, a breaking of $C_{4}$ symmetry is not required for this contribution to be finite. As already demonstrated in Ref.~\onlinecite{jiang2016}, also the symmetry-breaking contribution can obtain a finite expectation value bringing the system into a nematic phase.

To project out the symmetric contribution from $\chi^{\mu\nu}({\bf k},\sigma)$, we first note that under a $C_{4}$ rotation it transforms
as
\be 
\chi^{\mu\nu}({\bf k},\sigma) \to \mathcal{R}^{\mu\nu}[\chi] =
[\chi^{\prime}]^{\mu\nu}({\bf k}^{\prime},\sigma),
\ee
with
\be
[\chi^{\prime}]^{\mu\nu}({\bf k}^{\prime},\sigma) = \sum_{\mu^{\prime}\nu^{\prime}}
[R^{\mathrm{T}}]^{\mu\mu^{\prime}}
\chi^{\mu^{\prime}\nu^{\prime}}(M {\bf k},\sigma)
R^{\nu^{\prime}\nu},
\ee
where $R^{\mu\nu}$ are the elements of the representation matrix of a $C_{4}$ transformation
acting on the orbital degrees of freedom. The matrix $M$ on the other hand corresponds to the inverse
transformation, since momenta and real-space or orbital degrees of freedom transform oppositely.
The matrix $R$ acting on orbital degrees of freedom reads as
\be 
R =
\begin{pmatrix}
0 & 1 & 0 & 0 & 0 \\
-1 & 0 & 0 & 0 & 0 \\
 0 & 0 & -1 & 0 & 0 \\
 0 & 0 & 0 & -1 & 0 \\
 0 & 0 & 0 & 0 & 1 \\
\end{pmatrix},
\ee
while the  matrix $M$ acting on the Bloch vector reads
\be 
M =
\begin{pmatrix}
0 & -1 \\
1 & 0
\end{pmatrix}.
\ee
The invariant contribution can now be defined as (we note that $\mathcal{R}^{4} = \mathbbm{1}$)
\be
\chi_{\mathrm{br}} = \frac{1}{4}\left(\chi + \mathcal{R}[\chi] + \mathcal{R}^{2}[\chi] + \mathcal{R}^{3}[\chi]\right),
\ee
where we omitted matrix indices and momentum labels for brevity and $\mathcal{R}^{n}$ denotes
$\mathcal{R}$ applied $n$ times. Accordingly, the symmetry-breaking part is
\be 
\chi_{\mathrm{sb}} = \chi - \chi_{\mathrm{br}}.
\ee
Following Ref.~\onlinecite{jiang2016} we can further expand $ \chi^{\mu\nu}({\bf k},\sigma) $ in NN form factors,
\be 
f_{s}({\bf k}) & = & \cos(k_{x}) + \cos(k_y), \\
f_{d}({\bf k}) & = & \cos(k_{x}) - \cos(k_{y}), \\
f_{p_{x}}({\bf k}) & = & \sqrt{2} \mathrm{i} \sin(k_{x}), \\
f_{p_{y}}({\bf k}) & = & \sqrt{2} \mathrm{i} \sin(k_{y}),
\ee
as
\be
\chi^{\mu\nu}({\bf k},\sigma) = \sum_{A}\chi_{A}^{\mu\nu}(\sigma)f_{A}({\bf k}), \, A = s, d, p_{x}, p_{y}.
\ee
This decomposition of course carries over to $\chi_{\mathrm{br}}^{\mu\nu}({\bf k},\sigma)$ and $\chi_{\mathrm{sb}}^{\mu\nu}({\bf k},\sigma)$. We therefore have to determine the matrices $n_{0}^{\mu\nu}$, $M^{\mu\nu}$ as well as $\chi_{\mathrm{br},A}^{\mu\nu}(\sigma)$ and $\chi_{\mathrm{sb},A}^{\mu\nu}(\sigma)$, $A = s, p_{x}, p_{y}, d$ self-consistently within our mean-field approach. The components of $\chi_{\mathrm{br},A}^{\mu\nu}(\sigma)$ lead to a self-consistent renormalization of the hopping parameters, while the components of $\chi_{\mathrm{sb},A}^{\mu\nu}(\sigma)$ serve as nematic order parameters. As argued in Ref.~\onlinecite{jiang2016}, the coupling strength of $\chi_{\mathrm{br},A}^{\mu\nu}(\sigma)$ and $\chi_{\mathrm{sb},A}^{\mu\nu}(\sigma)$, respectively, need not be identical as they obey different symmetries and can in principle renormalize differently under the systematic elimination of high-energy excitations. We now denote the coupling of the symmetry-preserving part as $\tilde{V}$, while the coupling of the nematic part is now denoted as $\tilde{V}_{0}$ and in general $\tilde{V} \neq \tilde{V}_{0}$. One needs $\tilde{V}_{0} > \tilde{V}$ to produce a sizeable splitting of the electronic spectrum at the $M$ point in the 2-Fe BZ. The bond-order contribution to the Hamiltonian becomes with this replacement
\begin{eqnarray}
h_{\mathrm{BO}}^{\mu\nu}({\bf k},\sigma) & \to  &\tilde{h}_{\mathrm{BO}}^{\mu\nu}({\bf k},\sigma) =
-\frac{\tilde{V}}{2}
\begin{pmatrix}
 \chi_{\mathrm{br}}^{\mu\nu}({\bf k},\sigma) & 0  \\
0 & \chi_{\mathrm{br}}^{\mu\nu}({\bf k}+{\bf Q},\sigma) \\
\end{pmatrix} 
-\frac{\tilde{V}_{0}}{2}
\begin{pmatrix}
 \chi_{\mathrm{sb}}^{\mu\nu}({\bf k},\sigma) & 0  \\
0 & \chi_{\mathrm{sb}}^{\mu\nu}({\bf k}+{\bf Q},\sigma) \\
\end{pmatrix} 
+ (\mu \leftrightarrow \nu)^{\ast}.
\end{eqnarray}
Below we collect the matrices $\chi_{\mathrm{br},A}^{\mu\nu}(\sigma)$ and $\chi_{\mathrm{sb},A}^{\mu\nu}(\sigma)$, $A = s, p_{x}, p_{y}, d$, for the symmetry-preserving and symmetry-breaking contributions to the bond-order mean-fields, where we suppress the spin label for simplicity. Following Ref.~\onlinecite{jiang2016} we neglect contributions from $\chi_{A}^{12}$, $\chi_{A}^{21}$, $\chi_{A}^{34}$, $\chi_{A}^{43}$ and $\chi_{A}^{45}$, $\chi_{A}^{54}$ that are not compatible with the glide-plane symmetry. The coefficient matrices for the symmetry-preserving contribution read
\be 
\chi_{\mathrm{br},s} = 
\begin{pmatrix} 
\frac{1}{2}\left( \chi_{s}^{11}+ \chi_{s}^{22} \right) & 0 & 0 & 0 & 0 \\
0 & \frac{1}{2}\left(\chi_{s}^{11} + \chi_{s}^{22} \right) & 0 & 0 & 0 \\
0 & 0 & \chi_{s}^{33}  & 0 & 0 \\
0 & 0 & 0 & \chi_{s}^{44}  & 0 \\
0 & 0 & 0 & 0 & \chi_{s}^{55}   
\end{pmatrix},
\ee
\be 
\chi_{\mathrm{br},p_{x}}^{\mu\nu} = 
\begin{pmatrix} 
0 & 0 & \frac{1}{2}\left( \chi_{p_{x}}^{13} + \chi_{p_{y}}^{23} \right) & \frac{1}{2}\left( \chi_{p_{x}}^{14} + \chi_{p_{y}}^{24} \right)  & \frac{1}{2}\left( \chi_{p_{x}}^{15} - \chi_{p_{y}}^{25} \right)  \\
0  & 0  &  \frac{1}{2}\left( \chi_{p_{x}}^{23} - \chi_{p_{y}}^{13} \right) &  \frac{1}{2}\left( \chi_{p_{x}}^{24} - \chi_{p_{y}}^{14} \right) &  \frac{1}{2}\left( \chi_{p_{x}}^{25} + \chi_{p_{y}}^{15} \right) \\
\frac{1}{2}\left( \chi_{p_{x}}^{31} + \chi_{p_{y}}^{32} \right) & \frac{1}{2}\left( \chi_{p_{x}}^{32} - \chi_{p_{y}}^{31} \right) & 0 & 0 & 0 \\
\frac{1}{2}\left( \chi_{p_{x}}^{41} + \chi_{p_{y}}^{42} \right) & \frac{1}{2}\left( \chi_{p_{x}}^{42} - \chi_{p_{y}}^{41} \right) & 0 & 0 & 0 \\
\frac{1}{2}\left( \chi_{p_{x}}^{51} - \chi_{p_{y}}^{52} \right) & \frac{1}{2}\left( \chi_{p_{x}}^{52} + \chi_{p_{y}}^{51} \right) & 0 & 0 & 0  
\end{pmatrix},
\ee
\be 
\chi_{\mathrm{br},p_{y}}^{\mu\nu} = 
\begin{pmatrix} 
0 & 0 & -\frac{1}{2}\left( \chi_{p_{x}}^{23} - \chi_{p_{y}}^{13} \right) & -\frac{1}{2}\left( \chi_{p_{x}}^{24} - \chi_{p_{y}}^{14} \right)  & \frac{1}{2}\left( \chi_{p_{x}}^{25} + \chi_{p_{y}}^{15} \right)  \\
0  & 0  &  \frac{1}{2}\left( \chi_{p_{x}}^{13} + \chi_{p_{y}}^{23} \right) &  \frac{1}{2}\left( \chi_{p_{x}}^{14} + \chi_{p_{y}}^{24} \right) &  -\frac{1}{2}\left( \chi_{p_{x}}^{15} - \chi_{p_{y}}^{25} \right) \\
-\frac{1}{2}\left( \chi_{p_{x}}^{32} - \chi_{p_{y}}^{31} \right) & \frac{1}{2}\left( \chi_{p_{x}}^{31} + \chi_{p_{y}}^{32} \right) & 0 & 0 & 0 \\
-\frac{1}{2}\left( \chi_{p_{x}}^{42} - \chi_{p_{y}}^{41} \right) & \frac{1}{2}\left( \chi_{p_{x}}^{41} + \chi_{p_{y}}^{42} \right) & 0 & 0 & 0 \\
\frac{1}{2}\left( \chi_{p_{x}}^{52} + \chi_{p_{y}}^{51} \right) & -\frac{1}{2}\left( \chi_{p_{x}}^{51} - \chi_{p_{y}}^{52} \right) & 0 & 0 & 0  
\end{pmatrix},
\ee
\be 
\chi_{\mathrm{br},d}^{\mu\nu} = 
\begin{pmatrix} 
\frac{1}{2}\left( \chi_{d}^{11} - \chi_{d}^{22} \right) & 0 &  & 0 & 0 \\
0 & -\frac{1}{2}\left( \chi_{d}^{11} - \chi_{d}^{22} \right)  & 0 & 0 & 0 \\
0 & 0 & 0 & 0 & \chi_{d}^{35} \\
0 & 0 & 0 & 0 & 0 \\
0 & 0 & \chi_{d}^{53} & 0 & 0  
\end{pmatrix}.
\ee
The coefficient matrices for the symmetry-breaking contribution read
\be 
\chi_{\mathrm{sb},s} = 
\begin{pmatrix} 
\frac{1}{2}\left( \chi_{s}^{11} - \chi_{s}^{22} \right) & 0 & \chi_{s}^{13} & \chi_{s}^{14} & \chi_{s}^{15} \\
0 & -\frac{1}{2}\left(\chi_{s}^{11} - \chi_{s}^{22} \right) & \chi_{s}^{23} & \chi_{s}^{24} & \chi_{s}^{25} \\
\chi_{s}^{31} & \chi_{s}^{32} & 0  & 0 & \chi_{s}^{35} \\
\chi_{s}^{41} & \chi_{s}^{42} & 0 & 0 & 0 \\
\chi_{s}^{51} & \chi_{s}^{52} & \chi_{s}^{53} & 0 & 0
\end{pmatrix},
\ee
\be 
\chi_{\mathrm{sb},p_{x}} = 
\begin{pmatrix}
\chi_{p_{x}}^{11} & 0 & \frac{1}{2}\left( \chi_{p_{x}}^{13} - \chi_{p_{y}}^{23} \right) & \frac{1}{2}\left( \chi_{p_{x}}^{14} - \chi_{p_{y}}^{24} \right)  & \frac{1}{2}\left( \chi_{p_{x}}^{15} + \chi_{p_{y}}^{25} \right)  \\
0  & \chi_{p_{x}}^{22}  &  \frac{1}{2}\left( \chi_{p_{x}}^{23} + \chi_{p_{y}}^{13} \right) &  \frac{1}{2}\left( \chi_{p_{x}}^{24} + \chi_{p_{y}}^{14} \right) &  \frac{1}{2}\left( \chi_{p_{x}}^{25} - \chi_{p_{y}}^{15} \right) \\
\frac{1}{2}\left( \chi_{p_{x}}^{31} - \chi_{p_{y}}^{32} \right) & \frac{1}{2}\left( \chi_{p_{x}}^{32} + \chi_{p_{y}}^{31} \right) & \chi_{p_{x}}^{33} & 0 & \chi_{p_{x}}^{35} \\
\frac{1}{2}\left( \chi_{p_{x}}^{41} - \chi_{p_{y}}^{42} \right) & \frac{1}{2}\left( \chi_{p_{x}}^{42} + \chi_{p_{y}}^{41} \right) & 0 & \chi_{p_{x}}^{44} & 0 \\
\frac{1}{2}\left( \chi_{p_{x}}^{51} + \chi_{p_{y}}^{52} \right) & \frac{1}{2}\left( \chi_{p_{x}}^{52} - \chi_{p_{y}}^{51} \right) & \chi_{p_{x}}^{52} & 0 & \chi_{p_{x}}^{55}  
\end{pmatrix}
\ee
\be 
\chi_{\mathrm{sb},p_{y}}^{\mu\nu} = 
\begin{pmatrix} 
\chi_{p_{y}}^{11} & 0 & \frac{1}{2}\left( \chi_{p_{x}}^{23} + \chi_{p_{y}}^{13} \right) & \frac{1}{2}\left( \chi_{p_{x}}^{24} + \chi_{p_{y}}^{14} \right)  & -\frac{1}{2}\left( \chi_{p_{x}}^{25} - \chi_{p_{y}}^{15} \right)  \\
0  & \chi_{p_{y}}^{22}  &  -\frac{1}{2}\left( \chi_{p_{x}}^{13} - \chi_{p_{y}}^{23} \right) &  -\frac{1}{2}\left( \chi_{p_{x}}^{14} - \chi_{p_{y}}^{24} \right) &  \frac{1}{2}\left( \chi_{p_{x}}^{15} + \chi_{p_{y}}^{25} \right) \\
\frac{1}{2}\left( \chi_{p_{x}}^{32} + \chi_{p_{y}}^{31} \right) & -\frac{1}{2}\left( \chi_{p_{x}}^{31} - \chi_{p_{y}}^{32} \right) & \chi_{p_{y}}^{33} & 0 & \chi_{p_{y}}^{35} \\
\frac{1}{2}\left( \chi_{p_{x}}^{42} + \chi_{p_{y}}^{41} \right) & -\frac{1}{2}\left( \chi_{p_{x}}^{41} - \chi_{p_{y}}^{42} \right) & 0 & \chi_{p_{y}}^{44} & 0 \\
-\frac{1}{2}\left( \chi_{p_{x}}^{52} - \chi_{p_{y}}^{51} \right) & \frac{1}{2}\left( \chi_{p_{x}}^{51} + \chi_{p_{y}}^{52} \right) & \chi_{p_{y}}^{53} & 0 & \chi_{p_{y}}^{55}  
\end{pmatrix},
\ee
\be 
\chi_{\mathrm{sb},d}^{\mu\nu} = 
\begin{pmatrix} 
\frac{1}{2}\left( \chi_{d}^{11} + \chi_{d}^{22} \right) & 0 & \chi_{d}^{13}  & \chi_{d}^{14} & \chi_{d}^{15} \\
0 & \frac{1}{2}\left( \chi_{d}^{11} + \chi_{d}^{22} \right)  & \chi_{d}^{23} & \chi_{d}^{24} & \chi_{d}^{25}  \\
\chi_{d}^{31} & \chi_{d}^{32}  & \chi_{d}^{33} & 0 & 0 \\
\chi_{d}^{41}  & \chi_{d}^{42} & 0 & \chi_{d}^{44} & 0 \\
\chi_{d}^{51}  & \chi_{d}^{52} & 0 & 0 & \chi_{d}^{55}  
\end{pmatrix}.
\ee

\section{Band renormalization, nematic order and SDW ordering tendency}
\label{app:fs}

In this appendix we provide additional information on the massive band renormalization due to the self-consistent field $\chi_{br}^{\mu\nu}$ driven by the NN Coulomb repulsion with strength $\tilde{V}$ and the susceptibility to the formation a nematic state, reproducing some of the results already obtained in Ref.~\onlinecite{jiang2016}. We obtain an important new result by uncovering an increased tendency toward stripe-SDW formation at the flank of the nematic dome for weakened Coulomb repulsion. 
%
%
We here describe in some detail the influence of the NN Coulomb repulsion on the Fermi surface as shown in Fig.~\ref{fig:fs} in the main text. We put the onsite interactions to zero, $ U = J = U^{\prime} = J^{\prime} = 0$ and also neglect the symmetry-breaking part of the NN Coulomb repulsion by putting $\tilde{V}_{0} = 0$.  In Fig.~\ref{fig:fs}(a) we show the Fermi surface of the tight-binding band-structure~\cite{ningning2014} in the 1-Fe Brillouin zone (BZ) for electron filling $n = 6$, featuring the typical Fermi surface topology obtained from DFT calculations for iron-pnictide and iron-chalcogenide materials. The electron pockets at $X$ and $Y$ as well as the two central hole pockets at $\Gamma$ feature mixed orbital character, while the hole pockets at $M$ are dominated by the $d_{xy}$ orbital. Setting $\tilde{V} = 0.74$ eV close to the value that was found to move a van Hove singularity onto the Fermi surface~\cite{jiang2016}, we show the $C_{4}$ symmetric Fermi surface of the strongly renormalized band in Fig.~\ref{fig:fs}(c). We observe that increasing the NN Coulomb repulsion results in shrinking both electron and hole pockets. At the same time, the ellipticity of the electron pockets changes drastically and results in Fermi surfaces elongated along the $\Gamma - X$ and $\Gamma - Y$ directions, respectively. The orbital character of the pockets, however, remains unchanged. The enhancement in the single-particle density of states makes the band electrons susceptible to the formation of a ${\bf q} = (0,0)$ instability. Letting $\tilde{V}_{0} \neq 0$ this ordering-tendency leads to the stabilization of a uniform, nematic bond order~\cite{jiang2016} state with dominant $d$-wave character. The order parameters corresponding to other symmetry channels are typically finite due to the broken $C_{4}$ symmetry but do not appear as independent instabilities. We show the Fermi surface in a self-consistently stabilized nematic state in Fig.~\ref{fig:fs}(d). Additionally, we demonstrate the band renormalization in the $C_{4}$ symmetric state in a narrow window of $\tilde{V}$-values in Fig.~\ref{fig:bands}. As $\tilde{V}$ increases, the $d_{yz}$ dominated electronic band is shifted through the Fermi level while both electron and hole pockets become progressively smaller. In the nematic state, the system actually remains metallic and features a Fermi surface with only $C_{2}$ symmetry and a deformation of central hole pockets around $\Gamma $ and the electron pockets at either $X$ or $Y$, see Fig.~\ref{fig:fs}(d).
%
\begin{figure}[ht!]
\includegraphics[width=0.8\textwidth]{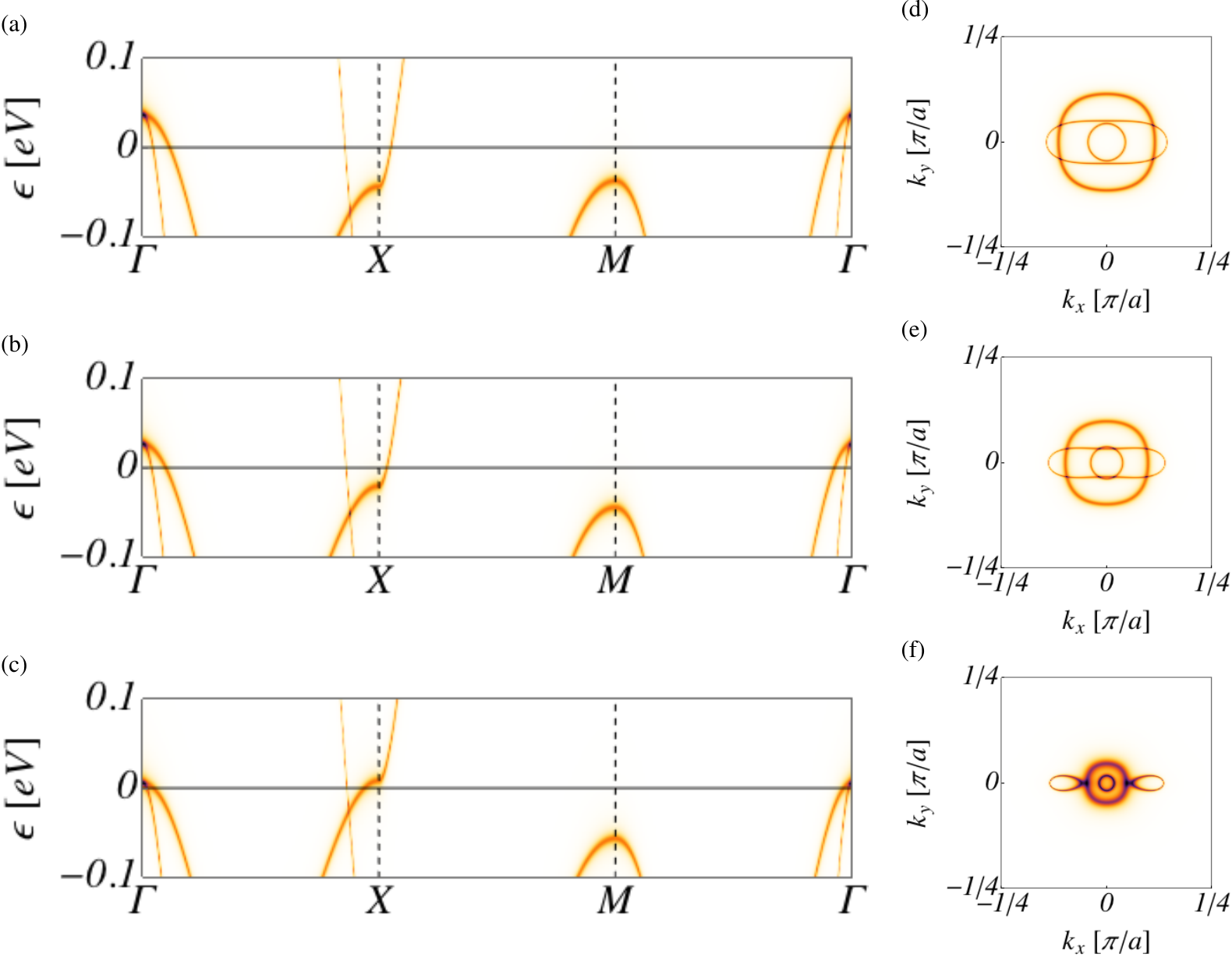}
\caption{(a) - (c) Electronic spectral weight along the high-symmetry cut $\Gamma-X-M-\Gamma$ in momentum space in a symmetric low-energy window of width $0.2\,\mathrm{eV}$ around the Fermi level. Here $ U, J = 0$ and $ \tilde{V}_{0} = 0$. As the interaction strength $\tilde{V}$ is increased from (a) $\tilde{V} = 0.69$ eV over (b) $\tilde{V} = 0.73 $ eV to (c) $\tilde{V} = 0.77 $ eV, both electron and hole pockets decrease in size. (d) - (f) The corresponding Fermi surfaces, where for clarity we folded the electron pockets around X onto the central hole pockets with the folding-vector $(\pi,0)$ to illustrate the varying degree of $(\pi,0)$ nesting. The same conclusions are obtained for the electron pocket at Y in the non-nematic phase. The nesting is close to optimal for (e). In the cases (d) and (f) nesting is in fact suppressed by matrix-element effects.
}
\label{fig:bands}
\end{figure}
%
As a next step we probe the tendency of the system to SDW formation in the renormalized $C_{4}$ symmetric phase. Previous theoretical studies\cite{mukherjee,kreisel} modelling NMR\cite{beak,imai09} and neutron scattering\cite{rahn,Zhao1,Zhao2} data have concluded that FeSe is close to a magnetic instability. We therefore compute the static spin susceptibility in the random phase approximation (RPA) in the transverse spin channel, defined by
\be 
& & \hspace{-1.50em} \chi^{\mathrm{RPA}}(\omega,{\bf q})|_{\omega \to 0}= \frac{1}{2\beta\mathcal{N}}\int_{0}^{\beta} \! d\tau \,
\sum_{\mu,\nu} \sigma_{\sigma_{1}\sigma_{2}}^{+}\sigma_{\sigma_{3}\sigma_{4}}^{-}\times \\
& & \sum_{{\bf k},{\bf k}^{\prime}} \langle \mathcal{T}_{\tau} 
c_{{\bf k} + {\bf q}\mu\sigma_{1}}^{\dagger}(\tau) c_{{\bf k}\mu\sigma_{2}}(\tau)
c_{{\bf k}^{\prime} - {\bf q}\nu\sigma_{3}}^{\dagger}(0) c_{{\bf k}^{\prime}\nu\sigma_{4}}(0)
\rangle_{\mathrm{RPA}}, \nn
\ee
where $\langle \cdots \rangle_{\mathrm{RPA}}$ refers the evaluation of the correlation function in the RPA approximation taking only the onsite interaction into account in the RPA-resummation process. We construct the bare propagator from the eigenstates of $h^{\mu\nu}({\bf k},\sigma)$, see \Eqref{eq:bloch}, and neglect the influence of the mean-fields $n_{0}^{\mu\nu}$ and $M^{\mu\nu}$ by setting $U,J = 0$. The influence of the band renormalization due to $\chi_{\mathrm{br}}^{\mu\nu}$ is kept, however. Here, we also introduced the fermionic operators in the imaginary time representation and the imaginary time-ordering operator $\mathcal{T}_{\tau}$ and introduced $\sigma^{+} = \sigma_{x} + i \sigma_{y}$ and $\sigma^{-} = \sigma_{x} - i \sigma_{y}$ with $\sigma_{x}, \sigma_{y}, \sigma_{z}$ denoting the Pauli matrices. A diverging static susceptibility points at the instability of the system to SDW formation with a particular ordering vector. In the following, we will restrict our focus to the ordering vector ${\bf q} = (\pi,0)$ or equivalently, by $C_{4}$ symmetry, ${\bf q} = (0,\pi)$. We have checked, however, that while small degree of incommensurability of the type $(\pi - \delta, \eta)$ with $|\eta|, |\delta| \ll \pi $ can in fact occur, in the $\tilde{V}$ range we are interested in, the SDW instability does not occur at, e.g., ${\bf q} = (\pi,\pi)$.
%
\begin{figure}[ht!]
\includegraphics[width=0.8\textwidth]{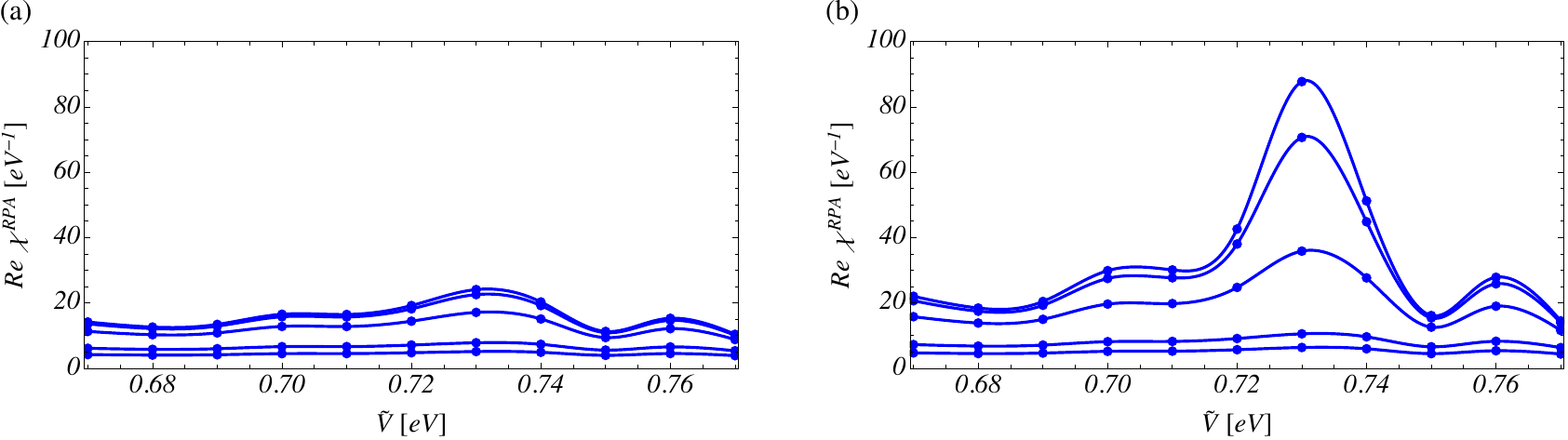}
\caption{The static RPA susceptibility $\chi^{\mathrm{RPA}}(\omega = 0,{\bf q})$ with momentum transfer ${\bf q} = (\pi,0)$
as a function of different interaction strength $\tilde{V}$ for (a) $U = 1.30 $ eV and (b) $U = 1.40 $ eV and increasing Hund's coupling (from bottom to top) $J = 0.1, 0.2, 0.3, 0.32, 0.324 \, \mathrm{eV} $ at temperature $T = 2$ meV. There is a clear enhancement of $(\pi,0)$ spin fluctuations in the vicinity of the point where the band renormalization pushes the van Hove singularity through the Fermi level.}
\label{fig:suscRPA}
\end{figure}
%
We focus on a range of the NN Coulomb repulsion $\tilde{V} \in [0.67,0.77]$ eV where the Fermi surface experiences a strong renormalization as shown in Fig.~\ref{fig:fs}(c). We further consider $U = 1.3, 1.4 $ eV for the onsite Hubbard-$U$ and vary the Hund's coupling for each of the cases independently. These parameters realize a SDW low-temperature state for $ \tilde{V} = 0 $ eV, i.e., for the unrenormalized band-structure. As shown in Fig.~\ref{fig:suscRPA}, we find that as $\tilde{V}$ is increased starting from $\tilde{V} = 0.67$ eV, the low-energy spin-fluctuations measured by $\chi^{\mathrm{RPA}}({\bf q})$ at the commensurate wave vector ${\bf q} = (\pi,0)$ increase and reach a maximum as a function of $\tilde{V}$ at $ \tilde{V}  \sim 0.73$ eV. Increasing the NN coupling $\tilde{V}$ further first leads to a decrease of low-energy spin fluctuations, but a second, sub-leading peak occurs at $\tilde{V} \sim 0.76$ eV. The renormalized band thus supports the formation of a SDW state with ordering vector ${\bf q} = (\pi,0)$ due to enhanced nesting and the proximity of the van Hove singularity for sufficiently large onsite interactions.

The dominant effect of allowing for a finite nematic order parameter $\Delta_{d}$ on the spin excitations is to promote $(\pi,0)$ fluctuations relative to $(0,\pi)$ fluctuations and vice versa, depending on which pair of electron pockets is pushed up or down due to the presence of nematic $d$-wave bond order. If $\Delta_{d} > 0 $ the $d_{yz}$ electron-band is pushed up and the corresponding pocket becomes smaller, while in the case $\Delta_{d} < 0$, it is the $d_{xz}$ electron-band that is pushed up. If we therefore assume that the nematic state sets in first as we decrease the temperature of the system, the nematic order selects the corresponding spin fluctuations and induces SDW order for appropriate values of the onsite interactions as the temperature is further decreased.

\end{widetext}

\end{document}